\documentclass[letterpaper, 10 pt, conference]{ieeeconf}  %
\pdfoutput=1 %

\IEEEoverridecommandlockouts                              %

\usepackage{graphics} %
\usepackage{epsfig} %
\usepackage{amsmath} %
\usepackage{amssymb}  %
\usepackage{leftidx}
\usepackage{multirow}
\usepackage[table,xcdraw]{xcolor}
\usepackage{subcaption}
\usepackage{algpseudocode}
\usepackage{algorithm}

\makeatletter
\let\NAT@parse\undefined
\makeatother
\usepackage[pagebackref=true,bookmarks=false,hidelinks]{hyperref}

\title{\LARGE \bf
Marginal Covariance of Parameters in New Observations
}

\author{Jianzhu Huai$^{1}$%
	\thanks{$^{1}$Jianzhu Huai was with CEGE, 
		The Ohio State University, Columbus, OH}%
}

\begin{document}

\maketitle
\thispagestyle{empty}
\pagestyle{empty}

\begin{abstract}
We have observed a common problem of solving for the marginal
covariance of parameters introduced in new observations. 
This problem arises in several situations, including augmenting parameters to a Kalman filter,
and computing weight for relative pose constraints.
To handle this problem, we derive a solution in a least squares sense.
The solution is applied to the above two instance situations and 
verified by independently reported results.
\end{abstract}

\section{Problem and Assumptions}
The problem is to estimate the covariance or information matrix of parameters $\mathbf{x}_n$ which 
is involved in new observations $\mathbf{z}$ of existing parameters 
$\mathbf{x}_m$ with prior knowledge of uncertainty.
The subscript \textit{m} denotes the dimension of the main parameters,
and the subscript \textit{n} denotes the dimension of the new parameters.

The observations as a vector $\mathbf{z}$ are related to parameters $\mathbf{x} = [\mathbf{x}_m^\intercal \enskip \mathbf{x}_n^\intercal]$ 
by observation coefficient matrix, 
$\mathbf{H} = [\mathbf{H}_m \enskip \mathbf{H}_n]$, \textit{i.e.},
\begin{equation}
\begin{split}
\mathbf{z} &= \mathbf{H} \mathbf{x} + \mathbf{w}\\
&=\begin{bmatrix}
\mathbf{H}_m & \mathbf{H}_n
\end{bmatrix} \begin{bmatrix}
\mathbf{x}_m \\
\mathbf{x}_n
\end{bmatrix} + \mathbf{w}
\end{split}
\label{eq:obs}
\end{equation}
The additive noise affecting observations, $\mathbf{w}$, is assumed to be Gaussian white noise, 
$\mathbf{w} \sim N(\mathbf{0}, \mathbf{R})$ where $\mathbf{R}$ is an diagonal matrix.
The coefficient matrices $\mathbf{H}_m$ and $\mathbf{H}_n$ could be constant or contingent on $\mathbf{x}_m$ or $\mathbf{x}_n$.

The prior knowledge about $\mathbf{x}_n$ is assumed to be captured by a
Gaussian distribution with mean $\hat{\mathbf{x}}_m$ and covariance $\mathbf{P}$.
That is, formally,
\begin{equation}
\begin{split}
\mathbf{y} &= \begin{bmatrix}
\mathbf{I}_m & 0_{m\times n}
\end{bmatrix} \begin{bmatrix}
\mathbf{x}_m \\ \mathbf{x}_{n}
\end{bmatrix} + \mathbf{v} \\
&= \mathbf{F}\mathbf{x} + \mathbf{v}
\end{split}
\label{eq:prior}
\end{equation}
where $\mathbf{v}$ is Gaussian white noise $\mathbf{v} \sim N(\mathbf{0}, \mathbf{P})$ and 
$\mathbf{F}$ is shorthand for the coefficient matrix.
$\mathbf{y}$ is the pseudo observation of $\mathbf{x}_m$,
and its realization takes the value of $\hat{\mathbf{x}}_m$.

\section{Solution}
Covariance for $\mathbf{x}_n$ can be derived following the standard least squares approach.
The below derivation is essentially the same as that for deriving the Kalman filter update \cite{jekeli_inertial_2001}.

The least squares estimation for $\mathbf{x} =[\mathbf{x}_m^\intercal \enskip \mathbf{x}_n^\intercal]^\intercal$ 
is given by minimizing the cost function $E$,
\begin{equation}
	E = (\mathbf{y} - \mathbf{F}\mathbf{x})^\intercal \mathbf{P}^{-1} (\mathbf{y} - \mathbf{F}\mathbf{x}) + (\mathbf{z} - \mathbf{H}\mathbf{x})^\intercal \mathbf{R}^{-1} (\mathbf{y} - \mathbf{H}\mathbf{x}).
\end{equation}
According to the least squares theory \cite[(3.13)]{schaffrin_notes_2019}, 
the information for the state $\mathbf{x}$ is given by the Hessian matrix $\boldsymbol{\Lambda}$,
\begin{equation}
\label{eq:info_mat}
\begin{split}
	\boldsymbol{\Lambda} &= \begin{bmatrix}
	\mathbf{F}^\intercal & \mathbf{H}^\intercal 
	\end{bmatrix} \begin{bmatrix}
	\mathbf{P} & \mathbf{0} \\
	0 & \mathbf{R}
	\end{bmatrix}^{-1}
	\begin{bmatrix}
	\mathbf{F} \\ \mathbf{H}
	\end{bmatrix} \\
	&= \begin{bmatrix}
	\mathbf{P}^{-1} + \mathbf{H}_m^\intercal \mathbf{R}^{-1} \mathbf{H}_m & \mathbf{H}_m^\intercal \mathbf{R}^{-1} \mathbf{H}_n \\
	\mathbf{H}_n^\intercal \mathbf{R}^{-1} \mathbf{H}_m & 
	\mathbf{H}_n^\intercal \mathbf{R}^{-1} \mathbf{H}_n
	\end{bmatrix}
	\end{split}
\end{equation}
which is inverse of the covariance of state estimate $\hat{\mathbf{x}}$, $\texttt{cov}(\mathbf{x})$.
To simplify notations, let's denote the blocks of $\boldsymbol{\Lambda}$ as 
\begin{equation}
\boldsymbol{\Lambda} = \begin{bmatrix}
\mathbf{A} & \mathbf{B} \\ \mathbf{C} & \mathbf{D}
\end{bmatrix}.
\label{eq:lamdba_shorthand}
\end{equation}

To obtain the marginal covariance for the new parameters $\mathbf{x}_n$, and their correlation to the main parameters $\mathbf{x}_m$,
we can use the block matrix inversion formulae \cite[(2.2), (2.3)]{lu_inverses_2002}
which obviously require nonsingular conditions.

A useful result about marginal covariance is obtained 
when both $\mathbf{D}$ and its Schur complement 
$\mathbf{A} - \mathbf{B}\mathbf{D}^{-1}\mathbf{C}$ are 
invertible (\textit{i.e.}, nonsingular). The entire covariance
for $[\mathbf{x}_m^\intercal \enskip \mathbf{x}_n^\intercal]^\intercal$, $\texttt{cov}(\mathbf{x})$
is given by 
\begin{equation}
\label{eq:cov_mat}
   	\left[\begin{array}{c|c}
   	(\mathbf{A}-\mathbf{B}\mathbf{D}^{-1}\mathbf{C})^{-1} & 
   	-(\mathbf{A}-\mathbf{B}\mathbf{D}^{-1}\mathbf{C})^{-1}
   	\mathbf{B}\mathbf{D}^{-1} \\
   	\hline
   	\begin{split}
   	-\mathbf{D}^{-1}\mathbf{C}\cdot \\
   	 (\mathbf{A}-\mathbf{B}\mathbf{D}^{-1}\mathbf{C})^{-1}
   	 \end{split} & 
   	\begin{split}
   	\mathbf{D}^{-1} + \mathbf{D}^{-1}\mathbf{C} \cdot \\
   	(\mathbf{A}-\mathbf{B}\mathbf{D}^{-1}\mathbf{C})^{-1}
   	\mathbf{B}\mathbf{D}^{-1}
   	\end{split}
   	\end{array}\right]
\end{equation}
whose bottom right block is the marginal covariance for $\mathbf{x}_n$.

In case that $\boldsymbol{\Lambda}$ is singular, it is practically less useful to obtain its inverse. 
Loosely speaking, this situation occurs because the new observations $\mathbf{z}$ 
do not fully constrain the new parameters $\mathbf{x}_n$.
In this case, it is often more appropriate and efficient to use the Cholesky factor of
 $\boldsymbol{\Lambda}$ (see \ref{subsec:rel_pose}).

The above results are obtained for linear observation models \eqref{eq:obs}, 
but they can be readily extended to nonlinear observation models.
In this case, the observation model is linearized at current state estimate
$\hat{\mathbf{x}} = [\hat{\mathbf{x}}_m^\intercal \enskip \hat{\mathbf{x}}_n^\intercal]^\intercal$, 
and the parameters in \eqref{eq:obs} and \eqref{eq:prior} are the perturbations in $\mathbf{x}$, 
\begin{equation}
	\delta \mathbf{x} = \begin{bmatrix}
	\delta \mathbf{x}_m \\ \delta \mathbf{x}_n
	\end{bmatrix}.
\end{equation}
These perturbations can be defined by a generalized $+$ operator, the $\boxplus$-operator \cite{hertzberg_integrating_2011},
in left- or right-multiplication manner.

\section{Applications}
Several problems arising from filtering-based visual inertial odometry
can be solved with the above results \eqref{eq:info_mat} and \eqref{eq:cov_mat}.
The following examines two problems: initializing covariance of landmark parameters in a EKF-SLAM method, 
and weighting relative pose constraints in pose graph optimization.

\subsection{Landmark Covariance Initialization}
For a point-based EKF-SLAM method (\textit{e.g.}, \cite{li_optimization_2013}) that includes landmark parameters in the state vector,
the covariance for landmark parameters and their correlation to the existing state need to be properly initialized 
as well observed landmarks are added to the state.

This initialization has been studied in \cite{li_optimization_2013}. 
To derive expressions for relevant covariance blocks, 
the authors assigned infinite covariance for landmark parameters in the first place, 
and then performed the Kalman filter update, 
and finally took the limits.
In contrast, our approach is less involved and more general while arriving at exactly the same results.

Assuming that a landmark is well observed (\textit{i.e.}, the required inverses exist),
the equivalency can be shown by multiplying the observation equation \eqref{eq:obs} with 
the $\mathbf{Q}$ matrix of QR decomposition of the observation matrix block $\mathbf{H}_n$ and then 
applying the result \eqref{eq:cov_mat} to the transformed observation equation.

We begin with writing out the linearized observation system for system error state $\delta \mathbf{x}$ 
and the error state $\delta \mathbf{f}$ of one landmark,
\begin{equation}
\mathbf{\delta z}
=\begin{bmatrix}
\mathbf{H}_x & \mathbf{H}_f
\end{bmatrix} \begin{bmatrix}
\delta\mathbf{x} \\
\delta\mathbf{f}
\end{bmatrix} + \mathbf{w}
\label{eq:linear_obs}
\end{equation}
where $\delta \mathbf{z}$ is the observation residual $\delta \mathbf{z} = \mathbf{z} - h(\mathbf{x})$ of dimension $k$, and 
the Jacobians of $h(\cdot)$ relative to $\delta \mathbf{x}$ and $\delta \mathbf{f}$ are $\mathbf{H}_x$ of 
size $k\times m$ and $\mathbf{H}_f$ of size $k \times 3$, respectively.
The linearization requires an estimate of landmark parameters which 
can be obtained by DLT \cite{hartley_multiple_2003} or Gauss-Newton optimization \cite{mourikis_multi_2007}.

The multiplication step separates the observations into two subsystems, one independent of $\mathbf{f}$,
and the other dependent on $\mathbf{f}$,
thanks to the QR decomposition of $\mathbf{H}_f$,
\begin{equation}
	\mathbf{H}_f = \mathbf{QR} = 
	\begin{bmatrix}
	\mathbf{Q}_c & \mathbf{Q}_o
	\end{bmatrix}\begin{bmatrix}
	\mathbf{R}_{c} \\ \mathbf{0}
	\end{bmatrix}
\end{equation}
where $\mathbf{Q}$ is an orthogonal matrix consisting of the 
column space basis $\mathbf{Q}_c$ of size $k\times 3$ and the left nullspace basis $\mathbf{Q}_o$ of size $k \times (k-3)$, 
and $\mathbf{R}_c$ is a $3\times 3$ upper triangular matrix.
Here we employs the assumption that the landmark is well observed.
Otherwise, $\mathbf{Q}_c$ will have less than 3 columns.

By left multiplying $\mathbf{Q}^\intercal$ to the observation system in \eqref{eq:linear_obs}, 
the two subsystems become obvious as below,
\begin{equation}
\label{eq:obs_nullspace}
\begin{split}
\begin{bmatrix}
\mathbf{Q}_c^\intercal \delta\mathbf{z} \\ \mathbf{Q}_o^\intercal \delta\mathbf{z}
\end{bmatrix}  &= 
\begin{bmatrix}
\mathbf{Q}_c^\intercal \mathbf{H}_x & \mathbf{Q}_c^\intercal \mathbf{H}_f
 \\ \mathbf{Q}_o^\intercal  \mathbf{H}_x & \mathbf{Q}_o^\intercal  \mathbf{H}_f
\end{bmatrix}
\begin{bmatrix}
\delta\mathbf{x} \\
\delta\mathbf{f}
\end{bmatrix} + 
\begin{bmatrix}
\mathbf{Q}_c^\intercal  \mathbf{w} \\ \mathbf{Q}_o^\intercal  \mathbf{w}
\end{bmatrix}
\\
\begin{bmatrix}
 \delta\mathbf{z}_c \\ \mathbf{\delta z}_o\end{bmatrix} &=\begin{bmatrix}
\mathbf{H}_{cx} & \mathbf{H}_{cf} \\
\mathbf{H}_{o} & 0
\end{bmatrix} \begin{bmatrix}
\delta \mathbf{x} \\ \delta \mathbf{f}
\end{bmatrix} + \begin{bmatrix}
\mathbf{w}_c \\ \mathbf{w}_o
\end{bmatrix} \\
\mathbf{w}_c &\sim N(\mathbf{0}, \mathbf{R}_c) \\
\mathbf{w}_o &\sim N(\mathbf{0}, \mathbf{R}_o)
\end{split}
\end{equation}
where the covariance for separated noise $\mathbf{w}_c$ and $\mathbf{w}_o$ are denoted by diagonal matrices
$\mathbf{R}_c$ of size $3\times 3$ and $\mathbf{R}_o$ of size $(k-3)\times (k-3)$,
and the projected Jacobians are $\mathbf{H}_{cx}$ of size $3\times m$, $\mathbf{H}_{cf}$ of size $3\times 3$,
and $\mathbf{H}_{o}$ of size $(k-3)\times m$.

The information matrix for the system can be written out according to \eqref{eq:info_mat} by plugging in the corresponding terms,
\begin{equation}
\label{eq:AB}
\begin{split}
\mathbf{A} &= 
\mathbf{P}^{-1} + \mathbf{H}_m^\intercal \mathbf{R}^{-1} \mathbf{H}_m \\
&= \mathbf{P}^{-1} + \begin{bmatrix}
\mathbf{H}_{cx}^\intercal & \mathbf{H}_{o}^\intercal
\end{bmatrix}
\begin{bmatrix}
\mathbf{R}_{c}^{-1} & \mathbf{0} \\ 
\mathbf{0} & \mathbf{R}_{o}^{-1}
\end{bmatrix}
\begin{bmatrix}
\mathbf{H}_{cx} \\ \mathbf{H}_{o}
\end{bmatrix} \\
&= \mathbf{P}^{-1} + \mathbf{H}_{cx}^\intercal \mathbf{R}_{c}^{-1} \mathbf{H}_{cx} + 
\mathbf{H}_{o}^\intercal \mathbf{R}_{o}^{-1} \mathbf{H}_{o} 
\\
\mathbf{B} &= \mathbf{H}_m^\intercal \mathbf{R}^{-1} \mathbf{H}_n \\
&= \begin{bmatrix}
\mathbf{H}_{cx}^\intercal & \mathbf{H}_{o}^\intercal
\end{bmatrix}
\begin{bmatrix}
\mathbf{R}_{c}^{-1} & \mathbf{0} \\ 
\mathbf{0} & \mathbf{R}_{o}^{-1}
\end{bmatrix}
\begin{bmatrix}
\mathbf{H}_{cf} \\ \mathbf{0}
\end{bmatrix} \\
&= \mathbf{H}_{cx}^\intercal \mathbf{R}_{c}^{-1} \mathbf{H}_{cf}
\end{split}
\end{equation}

\begin{equation}
\label{eq:CD}
\begin{split}
\mathbf{C} &= \mathbf{H}_n^\intercal \mathbf{R}^{-1} \mathbf{H}_m \\
&= \mathbf{H}_{cf}^\intercal \mathbf{R}_{c}^{-1} \mathbf{H}_{cx} \\
&= \mathbf{B}^\intercal
\\
\mathbf{D} &= \mathbf{H}_n^\intercal \mathbf{R}^{-1} \mathbf{H}_n \\
&= \begin{bmatrix}
\mathbf{H}_{cf}^\intercal & \mathbf{0}
\end{bmatrix}
\begin{bmatrix}
\mathbf{R}_{c}^{-1} & \mathbf{0} \\ 
\mathbf{0} & \mathbf{R}_{o}^{-1}
\end{bmatrix}
\begin{bmatrix}
\mathbf{H}_{cf} \\ \mathbf{0}
\end{bmatrix} \\
&= \mathbf{H}_{cf}^\intercal \mathbf{R}_{c}^{-1} \mathbf{H}_{cf}
\end{split}
\end{equation}

By using the augmented covariance equation \eqref{eq:cov_mat}, the system covariance with the new landmark can be expanded out.
The resultant expressions are identical to those derived in \cite{li_optimization_2013}.
For a sanity check, we expand the marginal covariance for the new landmark parameters by substituting the terms of \eqref{eq:AB} and \eqref{eq:CD} into the bottom right block of \eqref{eq:cov_mat}.
First notice that the inverse of $\mathbf{H}_{cf}$ exists because we assume the landmark is well observed.
Then the central inverse component can be simplified as
\begin{equation}
\begin{split}
\mathbf{A}-\mathbf{B}\mathbf{D}^{-1}\mathbf{C} &=
\mathbf{P}^{-1} + \mathbf{H}_{cx}^\intercal \mathbf{R}_{c}^{-1} \mathbf{H}_{cx} + 
\mathbf{H}_{o}^\intercal \mathbf{R}_{o}^{-1} \mathbf{H}_{o} \\
& \quad - \mathbf{H}_{cx}^\intercal \mathbf{R}_{c}^{-1} \mathbf{H}_{cf} \cdot 
\mathbf{H}_{cf}^{-1} \mathbf{R}_{c} \mathbf{H}_{cf}^{-\intercal} \cdot \\
& \quad \mathbf{H}_{cf}^\intercal \mathbf{R}_{c}^{-1} \mathbf{H}_{cx} \\
& = \mathbf{P}^{-1} + \mathbf{H}_{o}^\intercal \mathbf{R}_{o}^{-1} \mathbf{H}_{o}.
\end{split}
\end{equation}
By the Sherman-Morrison-Woodbury-Schur formula \cite[(A.6a)]{schaffrin_notes_2019},
the inverse of the above expression can be converted as
\begin{equation}
\begin{split}
(\mathbf{P}^{-1} + \mathbf{H}_{o}^\intercal \mathbf{R}_{o}^{-1} \mathbf{H}_{o})^{-1} =\\
\mathbf{P} - \mathbf{P} \mathbf{H}_{o}^\intercal (\mathbf{H}_{o}
\mathbf{P} \mathbf{H}_{o}^\intercal + \mathbf{R}_{o})^{-1} \mathbf{H}_{o}
\mathbf{P}
\end{split}
\label{eq:sherman}
\end{equation}
Finally, the marginal covariance for the landmark (see \eqref{eq:cov_mat}) can be derived as
\begin{equation}
\begin{split}
\texttt{cov}(\delta\mathbf{f}) &= \mathbf{D}^{-1} +
\mathbf{D}^{-1}\mathbf{C}(\mathbf{A}-\mathbf{B}\mathbf{D}^{-1}\mathbf{C})^{-1}\mathbf{B}\mathbf{D}^{-1} \\
&= \mathbf{H}_{cf}^{-1} \mathbf{R}_{c} \mathbf{H}_{cf}^{-\intercal} + \mathbf{H}_{cf}^{-1} \mathbf{R}_{c} \mathbf{H}_{cf}^{-\intercal} \cdot \\
& \quad \mathbf{H}_{cf}^\intercal \mathbf{R}_{c}^{-1} \mathbf{H}_{cx} \cdot
(\mathbf{P}^{-1} + \mathbf{H}_{o}^\intercal \mathbf{R}_{o}^{-1} \mathbf{H}_{o})^{-1} \cdot \\
& \quad \mathbf{H}_{cx}^\intercal \mathbf{R}_{c}^{-1} \mathbf{H}_{cf} \cdot 
\mathbf{H}_{cf}^{-1} \mathbf{R}_{c} \mathbf{H}_{cf}^{-\intercal} \\
&= \mathbf{H}_{cf}^{-1} \mathbf{R}_{c} \mathbf{H}_{cf}^{-\intercal} + \mathbf{H}_{cf}^{-1} \mathbf{H}_{cx} \cdot \\
& \quad (\mathbf{P}^{-1} + \mathbf{H}_{o}^\intercal \mathbf{R}_{o}^{-1} \mathbf{H}_{o})^{-1} 
\mathbf{H}_{cx}^\intercal \mathbf{H}_{cf}^{-\intercal}
\end{split}
\label{eq:cov_lmk}
\end{equation}
In view of \eqref{eq:sherman}, 
it is obvious that the above expression is the same as \cite[(30)]{li_optimization_2013}.

One benefit of separating the observations into two groups is that expressions in \eqref{eq:cov_mat} can be written out analytically as \eqref{eq:cov_lmk}.
Another benefit is that the covariance calculation \eqref{eq:cov_mat} can be divided into two smaller steps,
one for augmenting new parameters to the covariance matrix, 
and the other for updating covariance with classic EKF, as done in \cite{geneva_openvins_2019}.
The first step uses the projected observations $\delta \mathbf{z}_c$ that depend on both the system state and landmark parameters, 
to initialize the covariance for the landmark parameters and their correlation to the system state.
The analytic expressions \cite[(21)-(22)]{geneva_openvins_2019} for these covariance blocks 
can be derived by \eqref{eq:cov_mat}.
The second step uses the projected observations $\delta \mathbf{z}_o$ that do not depend on landmark parameters,
to update the current system state which includes the just augmented landmark parameters.
The net effect of the two steps can be shown analytically to be identical to 
the above approach (see \eqref{eq:cov_lmk}) that uses all observations in a single step.

Because the above derivation has little bearing on the number of landmarks, 
it should be able to augment multiple landmarks into the covariance matrix at once.

\subsection{Relative Pose Uncertainty}
\label{subsec:rel_pose}
Many odometry algorithms use relative pose constraints in graph-based optimization which 
often need a uncertainty estimate in order to properly weight 
them in optimization \cite{westman_degeneracy_2019}.

One type of relative pose constraint is calculated from 3D landmark - 2D feature correspondences.
The relative pose estimate can be solved by a PnP algorithm, \textit{e.g.}, \cite{kneip_novel_2011}.
The information for the relative pose can be obtained from the Hessian matrix of the linearized observation system
if we assume that the 3D landmarks are free of noise.

Otherwise, if we have prior knowledge about the uncertainty of these 3D landmarks,
then we may use the proposed method to compute the information matrix for the relative pose constraint.
In this case, the existing state is the 3D landmarks $\mathbf{x}_f = [\mathbf{f}_1^\intercal, \mathbf{f}_2^\intercal, \dots, \mathbf{f}_l^\intercal]^\intercal$, 
and the new state is the relative pose, $\mathbf{T}$.
The image observations $\mathbf{z}$ of these landmarks can be linearized and written in terms of the error state 
$\delta\mathbf{x}_f = [\delta \mathbf{f}_1^\intercal, \delta \mathbf{f}_2^\intercal, \dots, 
\delta \mathbf{f}_l^\intercal]^\intercal$ and $\delta\mathbf{T}$,
\begin{equation}
\delta \mathbf{z} = \begin{bmatrix}
\mathbf{H}_f & \mathbf{H}_T
\end{bmatrix} \begin{bmatrix}
\delta \mathbf{x}_f \\
\delta \mathbf{T}
\end{bmatrix} + \mathbf{w}
\end{equation}
where $\mathbf{w}$ is Gaussian white noise, $\mathbf{w} \sim N(\mathbf{0}, \mathbf{R})$.
The prior uncertainties for these landmarks may for instance come from
a landmark-based filter for vision-aided odometry, e.g., \cite{li_optimization_2013}, and they typically takes the form \eqref{eq:prior}.
Then the information matrix for the
combined system state can be immediately obtained with \eqref{eq:info_mat}.
The information for $\delta \mathbf{T}$. \textit{i.e.},
inverse of its marginal covariance, is 
\begin{equation}
	\boldsymbol{\Lambda}_T = \mathbf{D} - \mathbf{C} \mathbf{A}^{-1} \mathbf{B}
\end{equation}
in terms of \eqref{eq:lamdba_shorthand}.
Because block $\mathbf{A}$ corresponds to landmark parameters which are well constrained by the prior,
it should be always invertible.

To properly weight this relative pose constraint in optimization, 
the square root information matrix $\mathbf{R}$ often comes handy
for normalizing (whitening) the measurement error $\delta \mathbf{z}$.
It can be calculated by the $\mathbf{LDL}^\intercal$ decomposition of $\boldsymbol{\Lambda}_T$ 
to handle a possibly singular $\boldsymbol{\Lambda}_T$ which
may crop up due to uninformative observations, \textit{e.g.}, \cite{westman_degeneracy_2019}.
Formally, the decomposition is given as
\begin{equation}
\begin{split}
\boldsymbol{\Lambda}_T &= \mathbf{P^\intercal LDL^\intercal P} 
= \mathbf{R}^\intercal \mathbf{R} \\
\mathbf{R} &= \mathbf{\sqrt{D} L^\intercal P}
\end{split}
\end{equation}
where $\mathbf{D}$ is a diagonal matrix, $\mathbf{L}$ a lower triangular matrix with a unit diagonal, and $\mathbf{P}$ a permutation matrix.
As such, $\mathbf{R}$ is not necessarily upper triangular.

\section{Conclusion}
We derive a solution for computing the covariance of extra
parameters involved in new observations of existing parameters with prior knowledge.
This solution can be viewed as a special case of the Kalman filter update,
but is more general as it works with degenerate information matrices.
The derived expressions apply to several problems in vision-aided odometry, 
including initializing covariance for landmark parameters,
and computing square root information for relative pose constraints.
For covariance initialization of landmark parameters,
the proposed method arrives at expressions exactly matching with independently reported results
\cite{li_optimization_2013,geneva_openvins_2019}, thus proving its validity.

\section*{ACKNOWLEDGMENT}
The author thanks Charles Toth for helpful remarks.

\bibliographystyle{IEEEtran}
\bibliography{ms}

\end{document}